\documentclass[prl,amssymb,superscriptaddress,showpacs,twocolumn,aps]{revtex4}
\usepackage{graphicx}
\usepackage{psfrag}
\usepackage{amsfonts,amsmath}
\usepackage{color}
\usepackage{amsbsy}
\usepackage{amssymb}
\usepackage{lineno}

\begin{document}

\title{Counterion desorption and globule-coil transition of a polyelectrolyte under tension}
\author{Ajay S. Panwar, Mark A. Kelly, Buddhapriya Chakrabarti, and  M. Muthukumar$^{*}$}
\affiliation{Department of Polymer Science and Engineering, University of Massachusetts, Amherst,
MA 01003.}
\date{\today}
\email{muthu@polysci.umass.edu}
\begin{abstract}
We explore the mechanical response of a single polyelectrolyte chain under tension in good and poor solvents using a combination of simulation and theory. In poor solvents, where the equilibrium state of the chain is a collapsed globule, we find that the chain undergoes a globule-coil transition, as the magnitude of the force is increased beyond a critical value. This transition, where the polymer size changes discontinuously from a small to a large value, is accompanied by release of bound counterions from the chain. We explain these results by adhering to a statistical mechanical theory of counter-ion condensation on flexible polyelectrolytes.
\end{abstract}

\maketitle
A polyelectrolyte (PE) is a charged polymer with ionizable side groups on its backbone that dissociate in polar solvents, resulting  in a macro-ion and oppositely charged counterions\cite{Oosawa:71}. In solution, the size of a polyelectrolyte is intricately coupled to its effective charge. The consequence of this coupling is most significant when a polyelectrolyte undergoes a globule-coil transition. Such a phase transition can be effected by either changing the solvent quality (temperature) or applying a tensile force. In this letter, we employ Langevin dynamics simulations to examine the elastic response of a flexible polyelectrolyte chain that undergoes a globule-coil transition under externally applied tensile forces. We explain the simulation results using a theoretical calculation based on a double minimization of the PE free energy~\cite{Muthu:04}.

Our work is motivated by single molecule force spectroscopy (SMFS) experiments on biologically occurring polyelectrolytes\cite{McCauley:06}, such as proteins and DNA, where AFM and optical tweezers are used to examine their elastic response. The elastic response obtained from these experiments provides insight into their constitutive behavior\cite{Perkins:95}, bonding interactions\cite{Vladescu:05}, and binding energies\cite{Mihailovic:06}. Although these experiments can examine conformational changes of biopolymers, the role of charges in such transitions cannot be experimentally determined. However, using simulations\cite{Maurice:99}$-$\cite{Varga:02}, we can track polymer conformations and calculate its degree of ionization simultaneously.

\begin{figure}
\includegraphics[width=6.7cm]{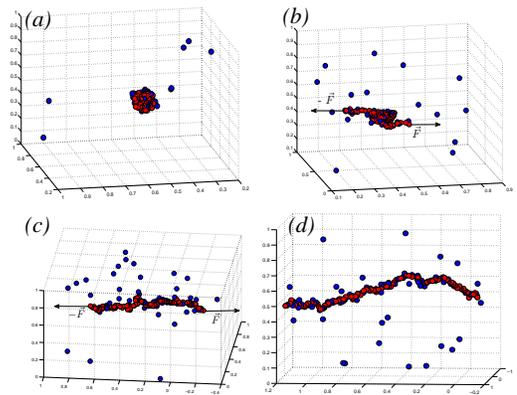}
\caption{Configuration snapshots of a single polyelectrolyte chain (red) under tension and counterions (blue). Panel (a) shows a PE globule with adsorbed counterions. The globular configuration just prior to the transition is shown in panel (b) while the coil configuration beyond a critical value of the applied force is shown in panel (c). Panel (d) shows a PE chain stretched to near complete extension at a higher value of applied force.}
\label{PEUnderTensionSchematic}
\end{figure}

In a poor solvent, a polyelectrolyte (PE) collapses into a globular configuration due to unfavorable solvent-polymer interactions. However, in order to minimize electrostatic repulsion between monomers, most of the counterions are adsorbed on the chain contour (Fig.~\ref{PEUnderTensionSchematic}(a)). In a good solvent, the PE adopts a swollen coil-like configuration because of favorable solvent-polymer interactions, and repulsive electrostatic interactions between monomers. The counterions now have the freedom to maximize their translational entropy, and some desorb from the chain to become mobile in solution. The fraction of mobile counterions, with respect to those adsorbed on the chain, depends on the relative strengths of electrostatic interactions and thermal fluctuations.

We simulate the stretching of a PE chain under equal and opposite tensile forces applied to its ends. Our main result is that at a critical value of applied force the PE chain undergoes a  globule-coil transition (Fig.\ref{PEUnderTensionSchematic}(b) and (c)), characterized by a discontinuous change in chain size. This transition is accompanied by a corresponding discontinuous increase in the degree of ionization (total number of free counterions in solution) of the chain. This is the first demonstration of counterion desorption accompanying a globule-coil transition, and calls for a careful analysis of previous theories where the ionization of the chain is assumed to be a parameter that remains constant throughout the transition~\cite{ConstantIonization}. We explain our simulation results using a variational formulation of the PE free energy that incorporates the stretching of the polymer under tension. Comparing the relative strengths of the different terms in the PE free energy, we find that at the transition, gain in mobile counterion entropy is greater than the chain-counterion adsorption energy and leads to counterion desorption.

The polyelectrolyte is modeled as a freely-jointed bead-spring chain comprised of $N = 100$ spherical beads connected by $N-1$ springs, where the length of each spring corresponds to a Kuhn step, $l_{0}$. The total charge $Q$ of the polyelectrolyte is uniformly distributed among all the beads (charge on each bead, $q = Q/N$). Charge neutrality is maintained by introducing $N$ monovalent counterions, modeled as spheres, each with a point unit electric charge $-q$. The polyelectrolyte and counterions are placed in a medium of uniform dielectric constant. The system is enclosed in a cubic box of edge length $50 l_{0}$ with periodic boundary conditions applied in each coordinate direction. Each bead represents a Brownian particle, whose motion is governed by the Langevin equation,
\begin{equation}
\label{eq:langevin}
m_{i} \frac{d^{2} {\bf r}_{i}}{d t^2} = - \zeta {\bf v}_{i} -\nabla_{i} U(r_{ij}) + {\bf F}^{r}_{i}(t) + {\bf F}^{ext}_{i},
\end{equation}
where $m_{i}$, ${\bf r}_{i}$ and ${\bf v}_{i}$ represent the mass, position and velocity of particle $i$, respectively. The random force ${\bf F}^{r}_{i}(t)$ acting on each bead satisfies the fluctuation-dissipation theorem, $\langle {\bf F}^{r}_{i}(t) {\bf F}^{r}_{j}(t^\prime) \rangle = 6 \zeta k_{B} T \delta(t - t^\prime) \delta_{ij}$, where $k_{B}$ is the Boltzmann constant, $T$ the absolute temperature, and $\zeta$ is the frictional drag. Externally applied forces, $F^{ext}_{i}$, of equal magnitude, $F$, but opposite directions act on the terminal beads of the chain. However, $F^{ext}_{i} = 0$ for the remaining polymer beads. Equation~\ref{eq:langevin} was solved numerically using the velocity-Verlet algorithm, with a time step $\Delta t = 0.001 \tau$, where $\tau = l_{0}(m / \epsilon_{LJ})^{1/2}$.

The net interaction potential, $U(r_{ij})$, is given by,
\begin{equation}
U(r_{ij}) = U_{LJ} + U_{bond} + U_{elec},\label{Potential-Eq}
\end{equation}
corresponding to excluded volume, bond stretching and electrostatic interactions, respectively. The excluded volume interaction between any two beads is described by,
\begin{eqnarray}
\label{eq:lj}
U_{LJ}(r_{ij}) = \left\{ \begin{array}{ll}
4\epsilon_{LJ}[{(\sigma/r_{ij})}^{12}-{(\sigma/r_{ij})}^{6}],
& r_{ij} \leq 2.5 \sigma, \\
0, \; r_{ij} > 2.5 \sigma,
\end{array} \right.
\end{eqnarray}
where $r_{ij} = | {\bf r}_{i} - {\bf r}_{j} |$, $\epsilon_{LJ}$ is the Lennard-Jones interaction parameter and $\sigma$ is the bead diameter. The bead diameters, $\sigma$, used in Eq.~(\ref{eq:lj}) for bead-bead, bead-counterion, and counterion-counterion interactions are 0.75 $l_0$, 0.55 $l_0$ and 0.35 $l_0$, respectively~\cite{Liu:02}. We use $\epsilon_{LJ}$, $l_0$ and $m$ as scales for energy, length, and mass, respectively. The masses of individual counterions and polymer beads are set to one half and a unit mass, respectively. The bonding interaction between adjacent beads is governed by the FENE potential given by,
\begin{equation}
U_{bond}(r_{b}) = - \frac{1}{2} k r_{b}^2 \ln \left[1 - \left(r_{b}/R_{max}\right)^2 \right],
\end{equation}
where $r_{b}$ is the separation distance between adjacent beads, $R_{max}=1.5l_{0}$ is the maximum allowable separation distance between bonded beads and $k$ is the spring constant. The electrostatic interactions between beads is given by the Coulombic potential
\begin{equation}
U_{elec}(r_{ij}) = \frac{q_{\rm i} q_{\rm j}}{4 \pi \epsilon_0 \epsilon_{r} r_{ij}},
\end{equation}
where $\epsilon_{0}$ is the vacuum permittivity and $\epsilon_{r}$ is the relative dielectric constant of the medium. The particle-particle particle-mesh (PPPM) method with a precision of 10$^{-3}$ was utilized to calculate the long-range electrostatic interactions in the system. Electrostatic interactions are parameterized by the Coulombic strength parameter $\Gamma = \frac{l_{B}}{l_0}$, where $l_{B}=\frac{e^2}{4 \pi \epsilon_{0} \epsilon_{r} k_{B} T}$ is the Bjerrum length. The simulations were executed in the open-source molecular dynamics simulation package LAMMPS\cite{Plimpton:95}.
\begin{figure}
\includegraphics[width=7cm]{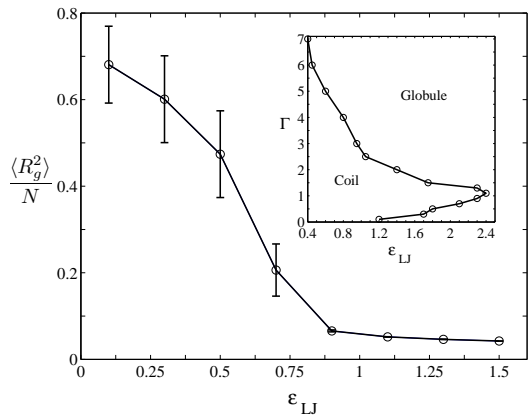}
\caption{The variation of radius of gyration of the polyelectrolyte chain $R_{G}$ as a function of the solvent quality $\epsilon_{\rm LJ}$ showing a coil to globule transition at $\epsilon_{\rm LJ} \approx 0.6$. The $\Gamma$ values were varied between $0.1$ and $7.0$, corresponding to low and high electrostatic strengths, respectively. Similarly, $\epsilon_{LJ}$ was varied between $0.1$ and $2.5$ corresponding to good and poor solvents, respectively. We compute the size of the chain, $R_g^2/N$, and its degree of ionization, $\alpha$, for these parameter ranges.}
\label{Phase-Diagram}
\end{figure}

Shown in Fig.~\ref{Phase-Diagram} is the variation of the normalized radius of gyration, $\langle R_g^{2}\rangle/N$, of the chain as a function of the solvent quality, $\epsilon_{LJ}$, at $\Gamma = 4.0$ (for bio-polymers, such as DNA, $\Gamma$ lies between $2.0$ and $4.0$\cite{Liu:02}). A coil to globule transition, signalled by a large change in $\langle R_g^{2}\rangle/N$ is observed with increasing $\epsilon_{LJ}$ (or decreasing solvent quality), with the transition occurring around $\epsilon_{LJ} \approx 0.6$. We note that for poor solvents $\epsilon_{LJ} \gtrsim 0.75$, the chain collapses into a globule of size smaller than Gaussian dimension. Inset shows the phase diagram in the $\Gamma-\epsilon_{LJ}$ plane, demarcating the globule and coil phases. The phase boundary is obtained by calculating $\langle R_g^{2}\rangle/N$ as a function of these parameters.

In order to extract the mechanical response of a single PE chain, it is subjected to a pulling and retraction force cycle, and its root-mean-square end-to-end distance is monitored as a function of time. The force is incrementally increased from $f = 0$ to $f = 10$ ($f = F/F_0$, where $F_0 = k_{B} T/l_0$, sets the scale of force) in the pulling cycle, and subsequently brought back to $f = 0$ in the retraction cycle. For a given value of applied force, the PE chain is equilibrated for $5 \times 10^6$ time steps. The normalized chain extension, $\langle x \rangle/N l_0$, is the time averaged root-mean-square end-to-end distance calculated over this duration. Simultaneously, the degree of ionization of the PE chain, defined as,
\begin{equation}
\label{eq:deg_ionization}
\alpha = \frac{ N - \left< n_{c} \right> }{N},
\end{equation}
where $\left< n_{c} \right>$ is the average number of counterions condensed on the polymer~\cite{Liu:02}, is also calculated over the force cycle.
\begin{figure}
\includegraphics[width=8.7cm]{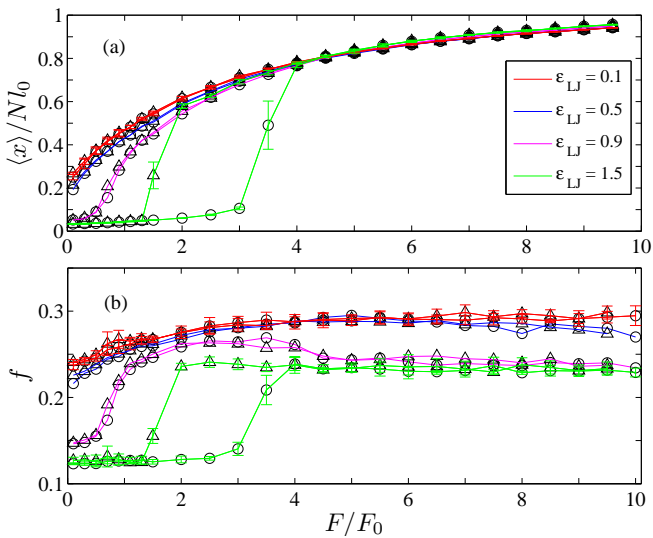}
\caption{Force vs.\ extension (top panel) and vs.\ degree of ionization (bottom panel), for $\Gamma = 4.0$, and as the solvent quality is varied during pulling (circles) and retraction (triangles). At zero force the chain is in a globular state for $\epsilon_{LJ} = 0.9$, and $1.5$, while it is in the coil phase for $\epsilon_{LJ} = 0.1,$ and $0.5$. The degree of ionization jumps from $\alpha \approx 0.12$ to $\alpha \approx 0.23$ at the coil-globule transition transition at $f_{c} \approx 3.0$ (pulling cycle) indicating counterion desorption.}
\label{Force-Extension-Result}
\end{figure}

Figure~\ref{Force-Extension-Result} shows the extension of the chain, $\langle x \rangle/Nl_0$, (Fig.~\ref{Force-Extension-Result}(a)), and its degree of ionization, $\alpha$,  (Fig.~\ref{Force-Extension-Result}(b)),
as a function of externally applied force, $f$ at $\Gamma = 4.0$ and $\epsilon_{LJ} = 0.1, 0.5, 0.9, 1.5$. For $\epsilon_{LJ} = 0.1, 0.5$, corresponding to a good solvent, $\left< x \right> / N l_0$ increases continuously from a swollen coil configuration to its maximum extent, as $f$ increases from $0.1$ to $10$. In both cases, the force-extension curves for pulling and retraction are identical. In contrast, in a poor solvent ($\epsilon_{LJ}$ = 1.5), the chain exists as a compact globule at $f = 0$ (see Fig.~\ref{PEUnderTensionSchematic}(a)). As $f$ is increased, $\left< x \right> / N l_0$ remains nearly constant until the force reaches a threshold value, $f_c \approx 3.0$. Beyond this value a large increase in $\left< x \right> / N l_0$ is observed for a very small increment in applied force (Figs.~\ref{PEUnderTensionSchematic}(b) and (c)), defining the force-induced globule-to-coil transition for a PE in a poor solvent.

A similar trend is observed in the variation of the degree of ionization of the PE (see Fig.~\ref{Force-Extension-Result}(b)) with the applied tensile force. For $\epsilon_{LJ} = 1.5$, $\alpha$ also shows a jump discontinuity at the same value of critical force, $f_c \approx 3.0$. The degree of ionization, $\alpha$, approximately doubles from a value of $0.12$ to $0.23$ at the transition. This suggests that the dramatic increase in the size of the chain at the transition leads to a reduced electrostatic repulsion between polymer segments. It is therefore favorable for the counterions to unbind from the chain and become mobile, thus gaining translational entropy that leads to a lowering of the system free energy.

We also note that both the force-extension curves show hysteresis for $\epsilon_{LJ} = 1.5$. This is a common feature of first order phase transitions where the system gets trapped in a metastable state, whose lifetime is longer than the observation time for simulations. The degree of ionization as a function of force shows similar hysteresis, indicating that the counterions relax on time scales much faster compared to the polymer relaxation time. Distinguishing long-lived metastable states from equilibrium phases is difficult, and requires a calculation of the free energies of the different phases.

\begin{figure}
\includegraphics[width=8.7cm]{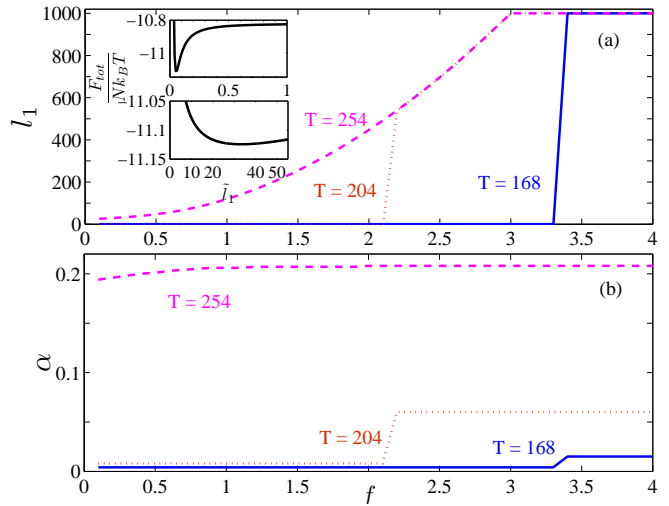}
\caption{Force vs.\ extension (a) and vs.\ degree of ionization (b) from variational calculation. Depending on temperature a globule-rod transition ($T = 168$, solid), globule-coil transition ($T = 204$, dotted) and a coil-rod transition ($T = 254$, dashed) is seen. Top inset shows the total free energy vs.\ extension, with minima corresponding to globule ($f = 0, \tilde{l}_{1} = 0.0002, \alpha = 0.0151$) (top) and coil ($f = 0.0003, \tilde{l}_{1} = 33, \alpha = 0.19$) (bottom) states at $T = 252$ K.}
\label{Force-Extension-Theory}
\end{figure}
A variational formulation of the free energy of a PE has been recently proposed\cite{Muthu:04} and analyzed in the context of globule-coil transition\cite{Kundagrami:08}. The total free energy of the PE system is given by,
\begin{equation}
F_{tot} = F_{cion} + F_{PE} + F_{ads} + F_{ten}. \label{Free-Energy-Theory}
\end{equation}
Counterion contribution to the free energy, represented by the first term,
\begin{eqnarray}
&&F_{c}/Nk_{B} T = \alpha \log \alpha + (1-\alpha)\log(1-\alpha) \nonumber \\ && + ( ( \alpha \rho + c_{s})  \log(\alpha \rho + c_{s}) + c_{s} \log c_{s} - ( \alpha \rho + 2 c_{s}) )/\rho \nonumber \\ && - \sqrt{4 \pi} \tilde{l}_{B}^{3/2} \left( \alpha \rho + 2 c_{s}\right)^{3/2}/\rho,\label{Counterion-Free-Energy}
\end{eqnarray}
where $\alpha$ is the degree of ionization of the chain, $c_s$ the salt concentration, $\rho$ the monomer density, and $\tilde{l}_{B}$ the Bjerrum length normalized by the Kuhn segment length $\tilde{l}_{B} = l_{B}/l_0$. The first two terms in Eq.\ref{Counterion-Free-Energy} correspond to the translational entropy of the adsorbed and mobile counterions respectively, while the third term is due to fluctuation of free ion density as given by Debye-H\"{u}ckel theory. The free energy of the polyelectrolyte, obtained from the Edwards Hamiltonian\cite{Muthu:04} by a variational calculation is given by
\begin{eqnarray}
\frac{F_{PE}}{N k_{B} T}&=&\frac{3}{2 N} \left(\tilde{l}_{1} - 1 - \log[\tilde{l}_{1}] \right) + \frac{4}{3}\left(\frac{3}{2 \pi} \right)^{3/2}\frac{w}{\tilde{l}^{3}_{1} \sqrt{N}} \nonumber \\ &+&
 \frac{w_{3}}{N \tilde{l}^{3}_{1}} + 2 \sqrt{\frac{6}{\pi}} \alpha^2 \tilde{l}_{B} \frac{\sqrt{N}}{\tilde{l}^{1/2}_{1}} \Theta_0(a), \label{Chain-Free-Energy-Eq}
\end{eqnarray}
where $\langle R^2 \rangle = N l^{2}_{0} \tilde{l_{1}}$, and $a \equiv \tilde{\kappa}^{2} N \tilde{l}_{1}/6$, $\kappa$ being the inverse Debye screening length given by $\tilde{\kappa}^{2} = 4 \pi \tilde{l}_{B} (\alpha \tilde{\rho} + 2 c_{s} )$. The last term in Eq.~\ref{Chain-Free-Energy-Eq} arises from the electrostatic interactions between segments of the polyelectrolyte chain and is given by $\Theta_{0}(a) = \frac{\sqrt{\pi}}{2} \left(\frac{2}{a^{5/2}}-\frac{1}{a^{3/2}} \right)\exp[a] {\rm erfc}[\sqrt{a}] + \frac{1}{3 a} +$ $ \frac{2}{a^2} - \frac{\sqrt{\pi}}{a^{5/2}} - \frac{\sqrt{\pi}}{2 a^{3/2}}$, while the first three terms correspond to configurational entropy of the chain, a repulsive excluded volume and steric three body interaction terms respectively.

Electrostatic interaction energy of ion-pairs formed due to counterions adsorbed on the chain is given by
\begin{equation}
F_{ads}/N k_{B} T = - ( 1 - \alpha) \delta \tilde{l}_{B},\label{Ion-Pair}
\end{equation}
where $\delta = (\epsilon/\epsilon_{l}) (l_0/d)$, is the ratio of the bulk dielectric constant $\epsilon$, to $\epsilon_l$ the local dielectric constant and $d$ is the length of the dipole. Contribution to the total free energy due to interactions between ion-pairs marginally affect the coil globule transition and are hence ignored.

In addition to these contributions the tensile force applied to the ends of the polyelectrolyte chain has a contribution,
\begin{equation}
F_{ten}/N k_{B} T = - f  \sqrt{\frac{\tilde{l}_{1}}{N}}, \label{Force}
\end{equation}
to the PE free energy\cite{footnote}. We obtain the chain extension ratio $\tilde{l}_{1}$, and the degree of ionization $\alpha$ as functions of force $f$ by a double minimisation of the total free energy $F_{tot}$.

Fig.~\ref{Force-Extension-Theory} summarizes our theoretical results. We consider a PE of $N = 1000$, at $c_s = 0$, $\rho = 0.0005$ and $\delta = 3.0$. The excluded volume interaction parameter is taken as, $w = 1 - \theta/T$. As an example, we consider $\theta = 400 K$ and $\tilde{l}_{B} = 900/T$. For the parameters used, a discontinuous globule-coil transition is observed for $w_3 < w^{*}_{3} = 8.5$\cite{Kundagrami:08}. In Fig.~\ref{Force-Extension-Theory}(a), we plot $\tilde{l}_{1}$ as a function of $f$ for $w_3 = 0.05$ and $T = 168, 204, 254$ K. At $T = 254$ K, the equilibrium coil state continuously stretches to a rod ($\tilde{l}_{1} = N$, used as an upper bound in our theory) with increasing $f$. At a lower temperature, $T = 204$ K, the globular PE undergoes a discontinuous globule-coil transition at a critical force, $f^{(1)}_{c} \approx 0.002$. On further increase in $f$, the coil stretches to a rod. A different discontinuous transition is observed for $T = 168$ K, where the globule transitions to a rod at a critical $f^{(2)}_{c} \approx 0.003$ bypassing the coil phase. This is because the coil phase is thermodynamically unstable for these temperatures\cite{Kundagrami:08}.

The PE free energy (at $T = 252$ K) as a function of $\tilde{l}_{1}$ is shown in Fig.~\ref{Force-Extension-Theory}(a) inset. At $f = 0$, the global free energy minimum occurs at $\tilde{l}_{1} \approx 0.002 $, corresponding to a globular PE (inset top panel). For $f > f_{c} = 0.0003$ the global minimum in free energy shifts to $\tilde{l}_{1} \approx 33$ corresponding to a coil phase. At this force, the limit of globule metastability is reached, leading to a discontinuous first-order globule-coil transition. The degree of ionization (Fig.~\ref{Force-Extension-Theory} (b)) follows a similar trend with the jump discontinuities appearing at the same value of $f$ at which the globule-coil transition occurs. We note that the trends of $\tilde{l}_{1}$ and $\alpha$ vs.\ $f$ are the same as those observed in simulations. A comparison of the terms in the total free energy suggests that $\alpha$ is coupled to $\tilde{l}_{1}$ through the electrostatic interaction term in Eq.~\ref{Chain-Free-Energy-Eq}. For $w < 0$, the two body term in Eq.~\ref{Chain-Free-Energy-Eq} tends to minimize $\tilde{l}_{1}$, and competes against the stretching term in Eq.~\ref{Force}. When $f > f_c$, $\tilde{l}_{1}$ increases abruptly to a high value and the electrostatic interaction term in Eq.~\ref{Chain-Free-Energy-Eq} becomes negligible. Then, $\alpha$ is determined by a competition between $F_{ads}$ and mobile counterion entropy.

In conclusion, we have explored the globule-coil transition in polyelectrolytes under mechanical tension. We have shown using Langevin dynamics simulations that at a critical value of the applied force, the globule-coil lengthening transition of the chain is accompanied by a correlated counterion desorption. In addition, our theoretical calculations indicate that the jump in the ionization of the chain at the globule-coil transition arises due to a gain in mobile counterion entropy that is greater than the chain-counterion adsorption energy. Our results can help understand conformational changes in biopolymers where electrostatic effects are important, and also aid interpretation of single molecule experiments involving polyelectrolytes.

The authors thank  grants from the NSF (DMR 0706454) and the NIH (5R01HG002776).


\begin{thebibliography}{99}
\bibitem{footnote} The mathematically exact term would have to be calculated by varying the exact end-to-end distribution function of a flexible chain with excluded volume
and electrostatic interactions with respect to $\sqrt{\langle R^2 \rangle}$, to be discussed in a future publication.
\end{thebibliography}
\end{document}